# IANEC : Investigation d'Archives Numériques d'Écrivains Contemporains

Travaux réalisés par le laboratoire GREYC

Emmanuel GIGUET

Université Caen Normandie, ENSICAEN, CNRS, Normandie Univ,
GREYC UMR 6072, F-14000 Caen, France
emmanuel.giguet@cnrs.fr

novembre 2025

## Table des matières

# Le projet IANEC

Le projet IANEC, Investigation d'Archives Numériques d'Écrivains Contemporains, porté par le Groupe de Recherche en Informatique, Image et Instrumentation de Caen (GREYC) et financé par le Ministère de la Culture dans le cadre de l'appel à projets Services Numériques Innovants (SNI), s'est donné pour objectif de développer des outils d'investigation numérique dédiés pour automatiser l'analyse des corpus d'archives de l'Institut mémoires de l'édition contemporaine (IMEC). Ce projet part du constat que les entrées d'archives numériques sont de de plus en plus fréquentes dans le contexte d'un centre d'archives contemporaines, et que les technologies numériques deviennent indispensables pour l'extraction, l'identification, le traitement et la description des corpus nativement numériques.

Le projet s'est déroulé en deux phases:
- • Phase 1 : La première partie du projet en 2024 a consisté à mettre en place les données, les outils logiciels côté GREYC et IMEC, et à recruter des étudiants en projet et stages sur financement IANEC. Quelques développements ont été réalisés dans cette première phase par trois stagiaires du GREYC. Une partie de ces développements, implémentés sur la plateforme G'DIP, a pu être testée par l'IMEC en 2024.
- Phase 2 : Au premier semestre 2025, 6 stagiaires du GREYC ont été recrutés. Des développements ont été réalisés tout au long du premier semestre, accompagnés côté IMEC par une stagiaire, pour suivre et tester les outils développés. Ces stagiaires étaient tous financés par le projet. Les outils et fonctionnalités développés en 2025 ont été volontairement rendus autonomes par rapport à la plateforme G'DIP pour faciliter l'interopérabilité et accélérer les phases de développement et de test dans le cadre du projet.

Le projet IANEC devait répondre aux besoins de l'IMEC selon deux axes de développement :

1. Un axe « Analyser les caractéristiques techniques de corpus nativement numériques » avec l'objectif d'obtenir un rapport technique exhaustif sur un corpus d'archives numériques donné, appliquant et enrichissant les fonctionnalités de la plateforme d'investigation numérique G'DIP développée par le GREYC.
2. Un axe « Analyser le contenu des contenus des documents », pour reconnaître et générer automatiquement certaines informations utiles à la description archivistique.

Ces travaux, coordonnés au sein du laboratoire GREYC par Christophe Rosenberger, professeur des universités à l'ENSICAEN, Emmanuel Giguet, chargé de recherche CNRS, et Tanguy Gernot, ingénieur de recherche CNRS, tous trois membres du thème Investigation numérique de l'équipe de recherche en cybersécurité SAFE, sont présentés dans les deux sections suivantes, et sont suivis d'un bilan.

# Travaux relatifs à l'analyse technique des corpus nativement numériques

## 1. Séparation des fichiers systèmes et applications des fichiers utilisateurs à l'aide de bases d'empreintes

Dans une image-disque, les fichiers systèmes et ceux issus d'applications représentent une part très importante du volume total de données. Leur présence en grande quantité complique le travail des archivistes, qui doivent passer beaucoup de temps à les identifier et les écarter afin de se concentrer sur les fichiers réellement produits par les utilisateurs.

Dans le projet IANEC, nous nous sommes donnés pour objectif d'automatiser l'identification des fichiers techniques présents dans les images-disques, afin de distinguer les fichiers systèmes et les fichiers issus d'applications de ceux réellement produits par les utilisateurs. Cette distinction est importante pour les archivistes de l'IMEC mais aussi et plus généralement pour tous les enquêteurs effectuant une investigation numérique : elle permet d'isoler rapidement les fichiers pertinents et de concentrer l'attention des spécialistes sur les documents personnels d'intérêt : courriers, documents contractuels, factures, photos, e-mails, par exemple.

Pour atteindre ce but, l'approche retenue a été de construire une base de données d'empreintes sur le modèle des Reference Data Set (RDS) du National Institute of Standards and Technology (NIST), mais adaptée aux besoins du projet. Le problème majeur est que les bases officielles du NIST ne couvrent pas les versions anciennes des systèmes et des applications Macintosh, qui sont pourtant centrales pour les collections traitées par les archivistes, notamment ceux de l'IMEC. Il a donc fallu compléter et enrichir ces bases par un travail spécifique sur les environnements Mac classiques. Les empreintes des fichiers des systèmes d'exploitation et des applications ont été extraites grâce à l'émulation des différentes versions de MacOS et à l'installation des applications dans ces environnements.

Concrètement, des outils spécifiques ont été développés pour extraire les empreintes d'images-disques ou de dossiers et les insérer automatiquement dans la base de données d'empreintes. Un soin particulier a été porté à la gestion des relations entre systèmes d'exploitation, applications et fichiers, afin de ne pas dupliquer inutilement des données et de garantir que les fichiers propres aux systèmes ne soient pas confondus avec ceux ajoutés par des applications. La base a été implantée dans PostgreSQL, plus adapté qu'un simple fichier SQLite grâce à son accès réseau, ses index et sa gestion collaborative.

Un outil graphique et en ligne de commande `hash_match_manager` a été développé pour permettre aux archivistes d'interagir facilement avec la base, et d'ajouter de nouvelles empreintes si nécessaire. Un plugin de la plateforme d'investigation numérique Autopsy a été réalisé afin d'intégrer cette fonctionnalité de séparation des fichiers d'intérêt dans un environnement forensique. Un outil autonome a été développé en parallèle et génère un rapport d'analyse interactif.

Au final, ce projet permet de disposer d'une base exploitable et extensible d'empreintes pour séparer les fichiers systèmes et les fichiers issus d'applications de ceux réellement produits par les utilisateurs. Des outils d'ajout automatisé de nouvelles empreintes ont été conçus. Deux interfaces permettent une mise en œuvre concrète de cette fonctionnalité d'identification des fichiers utilisateur sur les images fournies par l'IMEC. Si des limites subsistent, notamment sur la couverture des systèmes Mac, ce travail permet une exploitation concrète et collaborative des bases d'empreintes et laisse entrevoir un gain de temps substantiel dans le travail archivistique et plus généralement pour la communauté élargie de l'investigation numérique.

**Références :**

Titouan LE BRET - 4 mois. Investigation numérique d'archives d'écrivains contemporains : Analyse forensique et traitement d'archives numériques. Stage de M1 encadré par Tanguy Gernot, Emmanuel Giguet et Christophe Rosenberger. Du 3 mars au 4 juillet 2025. [Internship Report]

Guillaume HAUTOT - 4 mois. Investigation numérique d'archives d'écrivains contemporains. Stage de M1 encadré par Tanguy Gernot, Emmanuel Giguet et Christophe Rosenberger. Du 3 mars au 4 juillet 2025. [Internship Report]

**Illustrations :**

*Figure 1: Illustration du plugin Autopsy permettant d'identifier les fichiers systèmes et les fichiers d'applications*

## Rapport d'analyse de la source Macintosh_hd_leve.img

*Pour cette analyse, seuls ont été conservés les fichiers qui ne sont pas de taille nulle.*

### Informations concernant l'analyse :

- Date : 6 juin 2025 à 10h59-40 (Heure française)
- Informations sur la/les base(s) de données :
  - Base de données : **hashdb** | version : 2025.03.20 - GREYC-legacy-minimal
  - Base de données : **rds_legacy_minimal** | version : 2025.03.1 - legacy
  - Base de données : **hashdb_mac** | version : 2025.04.25 - GREYC-mac_os_6_to_10
- Nom de la data source : Macintosh_hd_leve.img
- Nom de l'examinateur : root

### Liste des fichiers sans correspondance :

| Nom du fichier | Taille | Hash MD5 |
|---|---|---|
| dmg/Desktop%20DF | 79554 | 9832c6dfbd563bbf6066a5b2dac0da36 |
| dmg/.fseventsd/000000000001e0cb | 70 | 7f3126d778be0d2ae446f0cea4d58d3d |
| dmg/Applications/ABBYY%20FineReader%205%20Sprint/Readme | 2392 | a6cb086c916d7427a13b0963cfdb6d1d |
| dmg/Desktop%20DB | 37376 | 560aefecb7ffd63f1cd4010335dab593 |
| dmg/.fseventsd/000000000001e0ce | 38 | ebb12e59f60c213f883d07e17ff3f2d5 |
| dmg/Applications/ABBYY%20FineReader%205%20Sprint/License | 9515 | a40304a939f1564f5f30c400f49c2135 |
| dmg/.fseventsd/000000000001e0cd | 71 | c266b7935d4cbdf591c7ae720a965e7f |
| dmg/.hotfiles.btree | 196608 | dd646dcc46b5a6d84713c9f7e6f08e0d |
| dmg/.DS_Store | 12292 | 613e303488a4f22167dc2075ead576eb |
| dmg/Applications/ABBYY%20FineReader%205%20Sprint/Launch%20FineReader%205%20Sprint | 175489 | 2fbc46371ee6550e8e5ab728ff335a8e |

« Précédent | Page 1 / 17898 | Suivant » | Lien vers JSON (117834 Ko)

### OS détectés par des correspondances sur la source :

| Nom de l'OS | Version | # occurrences | Base de données |
|---|---|---|---|
| Mac OS X Tiger | 10.4 | 131832 | hashdb |
| Mac OS 7 | 7.6.1 | 42 | hashdb |
| Mac OS X Tiger | 10.4 | 133655 | hashdb_mac |
| Mac OS X Panther | 10.3 | 55217 | hashdb_mac |
| Mac OS X Jaguar | 10.2 | 24806 | hashdb_mac |
| Mac OS 9 | 9.0.5 | 1073 | hashdb_mac |
| Mac OS 8 | 8.6 | 15 | hashdb_mac |
| Mac OS 8 | 8.5 | 15 | hashdb_mac |
| MacOS 8 | 8.0 | 15 | hashdb_mac |
| Mac OS 7 | 7.1 (m68k) | 10 | hashdb_mac |

### Correspondances avec des packages connus :

Correspondances détectées dans la base de données : hashdb

| ID | Nom du package | Version | Langue | OS | % de correspondance | # occurrences | Détails |
|---|---|---|---|---|---|---|---|
| 2 | Mac OS X Tiger(vierge) | 10.4 | French | Mac OS X Tiger | 78% | 126454 | Afficher/Masquer les fichiers |
| 5 | Adobe Acrobat | 6.0 | French | Mac OS X Tiger | 138% | 4142 | Afficher/Masquer les fichiers |
| 4 | StuffitDeluxe | 11 | French | Mac OS X Tiger | 11% | 188 | Afficher/Masquer les fichiers |
| 3 | Microsoft Office 2004 | 10.4 | French | Mac OS X Tiger | 63% | 1048 | Afficher/Masquer les fichiers |

*Figure 2: Illustration du rapport d'analyse au format HTML relatif à la présence de fichiers systèmes et fichiers d'application sur une image disque*

# 2. Lecture des fichiers sur partitions Mac et création d'un module FUSE

L'accès aux données de certains types de partitions utilisées par MacOS est souvent compliqué, ou du moins beaucoup moins simple que pour des formats de partitions plus classiques comme FAT, NTFS ou EXT. Cette difficulté concerne particulièrement les anciens systèmes de fichiers Macintosh (MFS, HFS, HFS+), qui conservent des spécificités comme la gestion des « resource forks » qui contiennent des métadonnées importantes pour l'investigation numérique.

Pour les archivistes, cette situation rend l'exploitation directe d'images-disques problématique, et conduit à recourir à des émulateurs spécifiques ou à des manipulations fastidieuses telles que la duplication des données vers un type de partitions facilitant l'investigation. L'objectif de ce travail était donc de fournir une solution permettant de lire facilement ces partitions sous Linux et de rendre directement accessibles toutes les informations qu'elles contiennent, y compris les resource forks, afin de simplifier et d'automatiser le traitement massif des images.

Pour cela, nous avons choisi de développer un module basé sur FUSE, ce qui permet de monter une image-disque dans l'espace utilisateur et de la rendre lisible comme n'importe quel autre système de fichiers classique. L'implémentation a été réalisée en C à partir de la bibliothèque libhfs. Le module expose les fichiers avec leurs data forks et leurs resource forks, ces derniers apparaissant dans un répertoire virtuel `.rsrc` parallèle à la structure normale, ce qui permet de conserver la logique du système de fichiers d'origine.

Le développement a nécessité la mise en place d'un ensemble de fonctions FUSE pour gérer l'ouverture, la lecture et la navigation dans l'arborescence. Des tests unitaires, l'utilisation de CMake et d'outils de vérification mémoire comme valgrind assure la robustesse du code. Le module fonctionne en lecture seule et permet d'accéder efficacement aux contenus des images Mac, mais certaines fonctionnalités, comme la gestion des liens symboliques propres au Mac ou le support du multithreading, restent limitées. Malgré ces points à améliorer, la création de ce module, baptisé `hfuse`, constitue un outil opérationnel et utile pour les archivistes, et plus généralement pour tout spécialiste de l'investigation, qui peuvent désormais parcourir et exploiter directement des images HFS/HFS+ dans leur environnement Linux.

**Référence :**

Guillaume HAUTOT - 4 mois. Investigation numérique d'archives d'écrivains contemporains. Stage de M1 encadré par Tanguy Gernot, Emmanuel Giguet et Christophe Rosenberger. Du 3 mars au 4 juillet 2025. [Internship Report]

# 3. Virtualisation des OS et interface d'émulation

L'exploitation des archives numériques de l'IMEC se heurte à de nombreuses difficultés techniques liées à l'obsolescence des environnements d'origine. Une partie des documents provient en effet de systèmes Macintosh des années 1990 à 2010. Ces environnements posent plusieurs problèmes : d'abord, l'obsolescence matérielle, car Mac OS Classic fonctionnait sur des architectures matérielles PowerPC ou M68k aujourd'hui disparues. Ensuite, l'obsolescence logicielle, puisque de nombreux formats de fichiers comme ClarisWorks, QuarkXPress ne sont plus directement lisibles avec les logiciels actuels. Enfin, des contraintes fortes d'interopérabilité subsistent, car même lorsqu'un fichier est lisible, il faut en garantir l'intégrité et la cohérence dans un environnement moderne, et sans perte de métadonnées.

Pour répondre à ces problèmes, le projet a poursuivi un double objectif. Le premier était de mettre en place des environnements d'émulation capables de reproduire fidèlement Mac OS quelles qu'en soit la version, afin de rendre accessibles les documents dans leur contexte logiciel d'origine. Le second objectif était de simplifier l'usage de ces environnements en concevant un outil logiciel doté d'une interface unifiée, masquant les difficultés liées aux différents systèmes d'émulation, afin que les archivistes puissent exploiter ces technologies sans avoir à manipuler directement des émulateurs complexes.

L'intérêt d'une telle approche est majeur pour la préservation numérique. L'émulation garantit l'accès aux documents dans des conditions proches de celles de leur production, ce qui permet de préserver la structure, l'apparence et parfois même l'expérience d'utilisation. L'interface développée, en normalisant et en automatisant des tâches comme la préparation de disques virtuels ou la configuration des machines, réduit les risques d'erreurs et rend la manipulation accessible à des utilisateurs non spécialistes. L'outil répond ainsi à un double enjeu : la conservation fidèle du patrimoine numérique et la praticité de son exploitation au quotidien.

Pour atteindre ces résultats, deux technologies principales ont été utilisées. SheepShaver a permis d'émuler efficacement Mac OS Classic, de la version 7.5.2 à 9.0.2, avec de bonnes performances et une communauté active. QEMU, plus complexe à configurer, a offert la possibilité d'émuler des environnements PowerMac G3/G4 et d'exécuter les premières versions de Mac OS X (jusqu'à 10.5). Les deux outils se sont révélés complémentaires et ont permis de couvrir l'ensemble des besoins. En parallèle, une interface en mode texte a été développée en Rust avec la bibliothèque Ratatui. Cette interface regroupe dans une même application les différents émulateurs, gère automatiquement les profils et les copies de disques, et propose une architecture modulaire permettant d'ajouter facilement de nouveaux environnements.

Les résultats obtenus sont concrets et directement exploitables. Il est désormais possible de lancer en quelques étapes des environnements Mac OS Classic ou Mac OS X préconfigurés, de charger des disques issus des archives et de les parcourir sans altérer les originaux. L'interface permet de travailler sur des copies, d'attacher de nouveaux supports et de dupliquer des environnements existants, tout en garantissant une compatibilité avec les infrastructures de l'IMEC. Malgré certaines limites (performances perfectibles de QEMU, impossibilité de démarrer directement sur certains disques originaux), le projet a abouti à un outil fonctionnel qui simplifie considérablement l'usage de l'émulation dans un contexte archivistique.

**Références :**

Hamid BEN OMAR - 4 mois. Intégration de modules d'investigation G'DIP à Autopsy et Préparatifs à l'investigation numérique sur d'anciens Mac OS. Stage de 2A encadré par Emmanuel Giguet et Tanguy Gernot. Du 15 avril au 19 août 2024. [Internship Report]

Matthias DAVID - 4 mois. Investigation numérique d'archives d'écrivains contemporains : Fournir un service numérique innovant à l'Institut mémoires de l'édition contemporaine. Stage de M1 encadré par Tanguy Gernot, Emmanuel Giguet et Christophe Rosenberger. Du 3 mars au 4 juillet 2025. [Internship Report]

**Illustrations :**

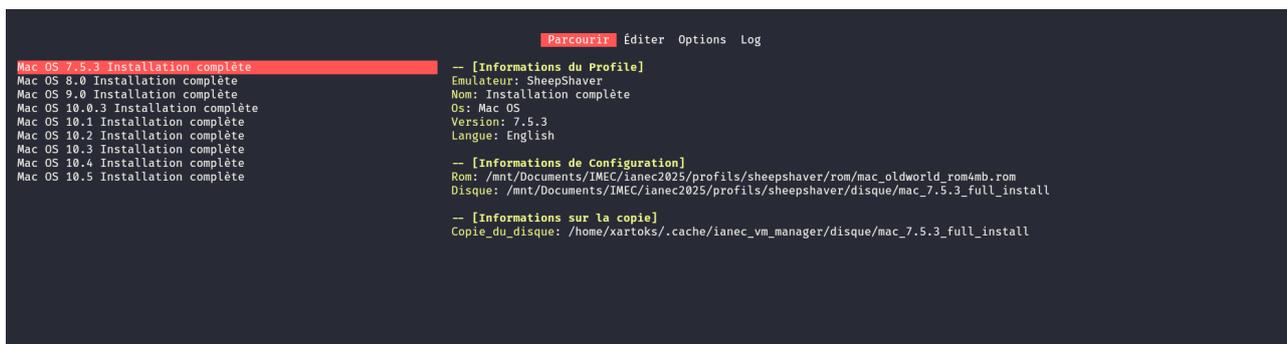

*Figure 3: Illustration de l'interface de virtualisation multi OS*

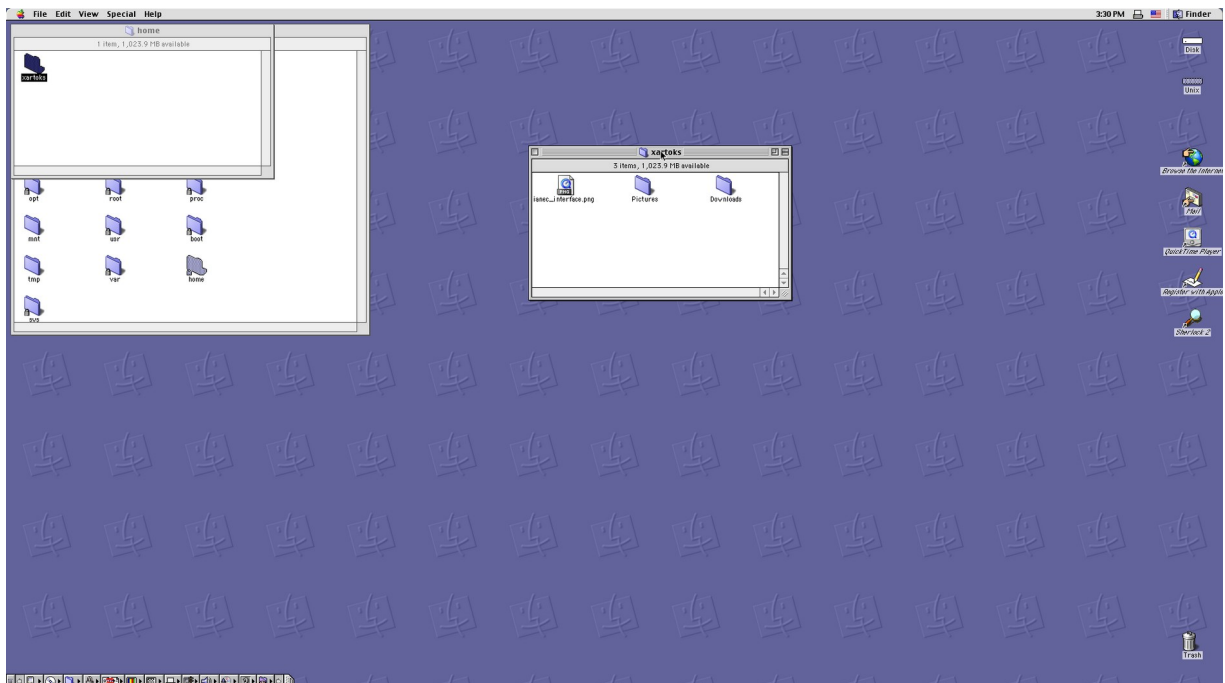

*Figure 4: Illustration de la virtualisation de MacOS 9 sous SheepShaver, lancé depuis l'interface de virtualisation multi OS*

## 4. Amélioration de l'interface de G'DIP

La plateforme d'investigation numérique G'DIP réalisée au sein du GREYC n'ayant jamais été utilisée en dehors du laboratoire, la collaboration avec l'IMEC a fait émerger la nécessité de repenser l'ergonomie de l'interface afin de faciliter sa prise en main. Tel a été l'objet de ce stage.

**Référence :**

Arthur LOMBARDO - 2 mois. Amélioration de l'interface utilisateur de la plateforme d'investigation numérique G'DIP. Stage de M1 encadré par Emmanuel Giguet et Tanguy Gernot. Du 27 mai au 19 juillet 2024.

# Travaux relatifs à l'analyse sémantique des documents numériques

## 5. Classification automatique des documents textuels

Les fonds d'archives numériques de l'IMEC contiennent une masse très hétérogène de documents textuels : correspondances privées, brouillons littéraires, journaux intimes, factures ou encore contrats. Cette diversité combinée à l'ampleur des fonds rend l'analyse manuelle longue, fastidieuse et difficile à étendre à l'ensemble des données à traiter.

L'objectif poursuivi par les travaux sur la classification automatique des documents textuels est de mettre en place des classifieurs capables d'automatiser l'organisation de ces documents selon des catégories pertinentes pour l'archivage : administratif, correspondance, activité d'écriture, vie privée, etc. Ces catégories ont été proposées par l'IMEC sous la forme d'une taxonomie. L'intérêt d'un système de classification automatique est de faciliter le travail des archivistes en organisant les documents personnels en fonction de leur type, avec comme conséquence indirecte de réduire les temps de traitement de chaque fonds.

Les approches mises en œuvre dans le cadre du projet reposent sur deux grands axes :

- Le machine learning classique, utilisant des représentations vectorielles des mots enrichies par des catégories lexicales pour entraîner des modèles supervisés classiques, type SVM, Random Forest, XGBoost.

- L'apprentissage par transfert et les grands modèles de langues, avec des modèles comme BART ou Mistral adaptés aux catégories archivistiques. Diverses expériences ont été menées telles que la spécialisation du modèle BART ou l'usage de Mistral pour explorer la génération de résumés ou de descriptions automatiques.

Une comparaison croisée a montré que les deux approches sont complémentaires : le machine learning est très fiable sur les structures lexicales explicites, tandis que les modèles de langage capturent mieux les nuances sémantiques. Les divergences apparaissent surtout sur les cas limites, où une vérification humaine reste nécessaire. Enfin, deux expériences complémentaires, le regroupement de documents (clustering) thématiquement similaires, et la recherche augmentée par génération (RAG) ont montré l'intérêt de combiner la classification automatique et la recherche sémantique, ouvrant la voie à des interfaces interactives pour l'exploration des fonds.

**Références :**

Dihia SLIMANA - 4 mois. Investigation numérique d'archives d'écrivains contemporains. Stage de M1 encadré par Tanguy Gernot, Emmanuel Giguet et Christophe Rosenberger. Du 3 mars au 4 juillet 2025. [Internship Report]

Mohammed Salah AISSAOUI - 4 mois. Investigation numérique d'archives d'écrivains contemporains. Stage de M1 encadré par Tanguy Gernot, Emmanuel Giguet et Christophe Rosenberger. Du 3 mars au 4 juillet 2025. [Internship Report]

**Illustration :**

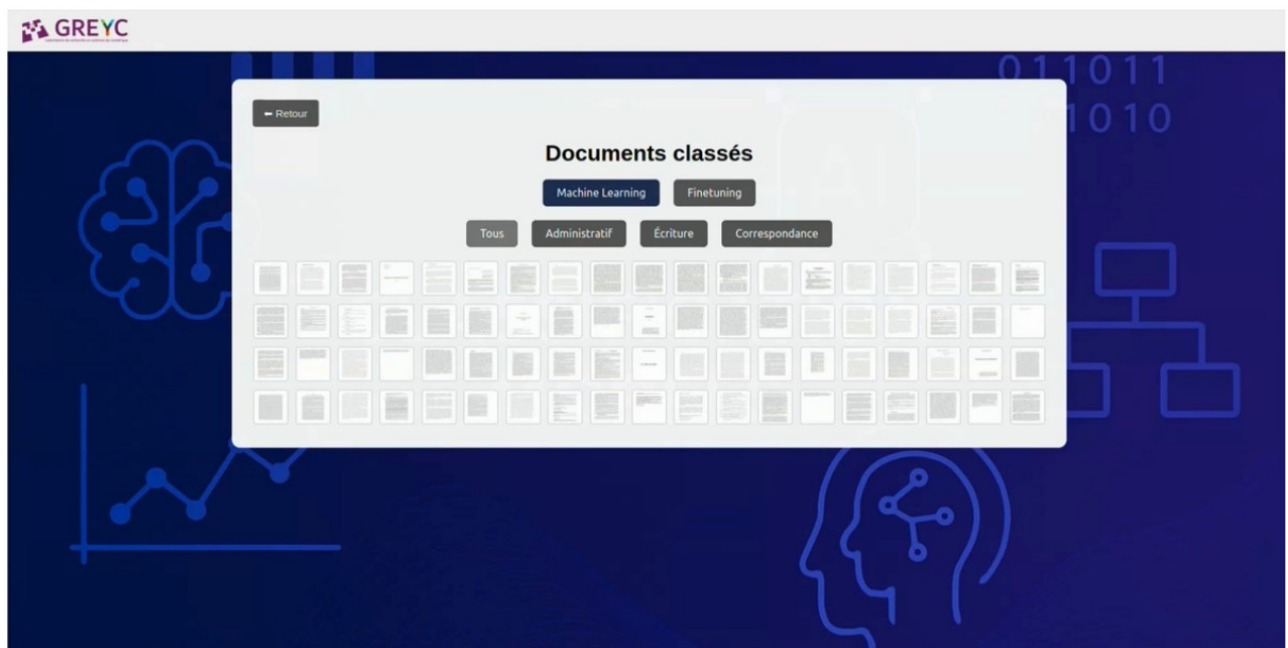

*Figure 5: Illustration de la plateforme de classification des documents textuels nativement numériques*

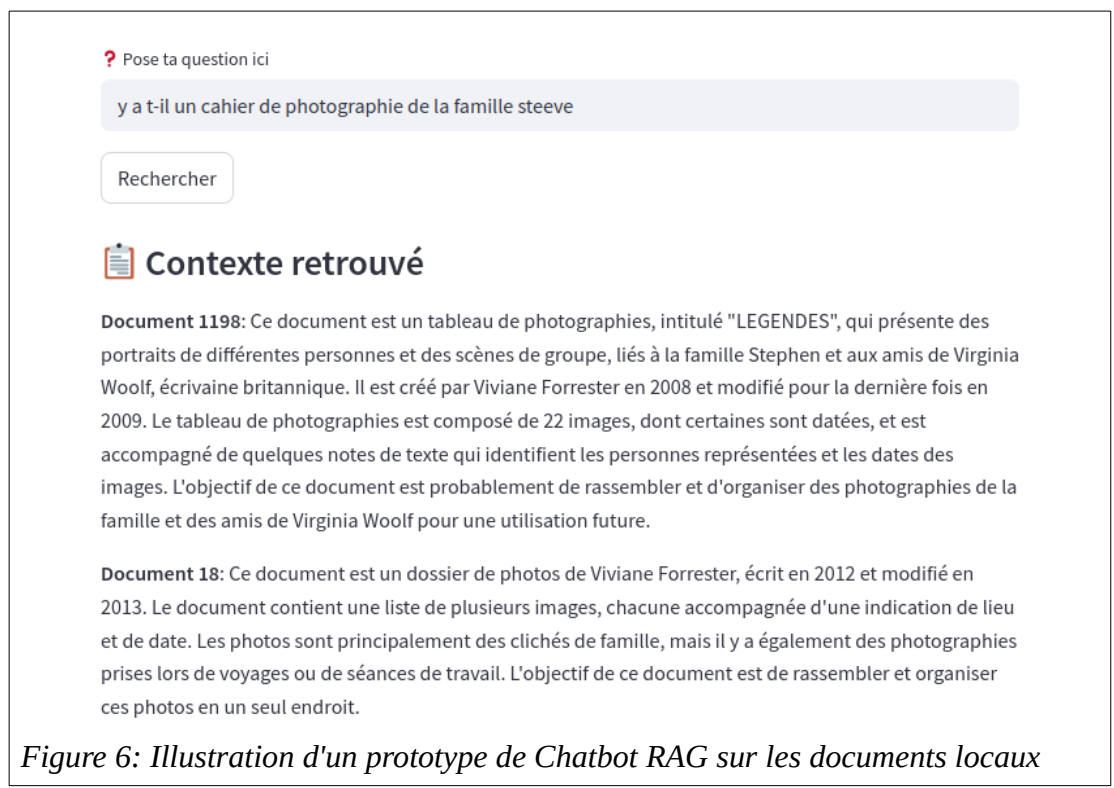

*Figure 6: Illustration d'un prototype de Chatbot RAG sur les documents locaux*

# 6. Analyse et classification des images (documents, photos, logos)

Les fonds d'archives numériques contiennent aussi une grande variété d'images : scans de documents imprimés ou manuscrits, photos personnelles, captures d'écran, icônes et logos. Ici encore, la difficulté principale réside dans l'hétérogénéité des fichiers, à laquelle s'ajoute une qualité souvent très variable des images.

L'objectif des travaux était donc double : d'une part, distinguer les grandes catégories visuelles (documents, photos de personnes, icônes/logos, autres images), d'autre part extraire le contenu textuel des images de documents pour les intégrer à la classification globale. L'intérêt est de ce travail est de mettre en avant les documents porteurs d'information, tout en filtrant les éléments accessoires ou décoratifs.

Deux grands outils d'extraction du contenu textuel provenant d'image de document ont été testés : un OCR classique, Tesseract, qui a rapidement montré ses limites pour le traitement de textes manuscrits, inclinés ou de mauvaise qualité. En outre, cet outil n'est pas conçu pour distinguer automatiquement les images pertinentes des images résiduelles sans intérêt archivistique. Il doit être suivi de traitements de plus haut niveau. Un outil innovant, Mistral-OCR a montré de bien meilleures performances, comparées à celles d Tesseract, notamment pour les manuscrits et les textes dégradés.

Plusieurs stratégies complémentaires de classification ont été testées :

- La classification visuelle par deep learning : des modèles préentraînés comme ResNet et surtout Vision Transformer (ViT) ont été adaptés par fine-tuning. Les résultats obtenus sont très prometteurs pour la classification en quatre classes (photo, document, icône, autre). Ils doivent cependant être validés sur des corpus autres.

- Les modèles multimodaux : PixTral-12B a été testé pour associer analyse visuelle et génération de descriptions, ce qui a permis d'améliorer la lisibilité des résultats et de limiter les erreurs liées aux hallucinations.

Les expérimentations ont donné naissance à des prototypes opérationnels permettant de charger des dossiers complets, de classer simultanément documents et images, de visualiser les prédictions et d'exporter les résultats. Ces outils offrent ainsi une chaîne cohérente pour trier, décrire et exploiter les fonds numériques.

**Références :**

Thomas VARIN - 3 mois. Détection et reconnaissance de visage dans des images d'archive. Stage de 2A encadré par Emmanuel Giguet et Tanguy Gernot. Du 3 juin au 23 août 2024. [Internship Report]

Dihia SLIMANA - 4 mois. Investigation numérique d'archives d'écrivains contemporains. Classification automatique de documents. Stage de M1 encadré par Tanguy Gernot, Emmanuel Giguet et Christophe Rosenberger. Du 3 mars au 4 juillet 2025. [Internship Report]

Mohammed Salah AISSAOUI - 4 mois. Investigation numérique d'archives d'écrivains contemporains. Classification automatique de documents. Stage de M1 encadré par Tanguy Gernot, Emmanuel Giguet et Christophe Rosenberger. Du 3 mars au 4 juillet 2025. [Internship Report]

**Illustrations :**

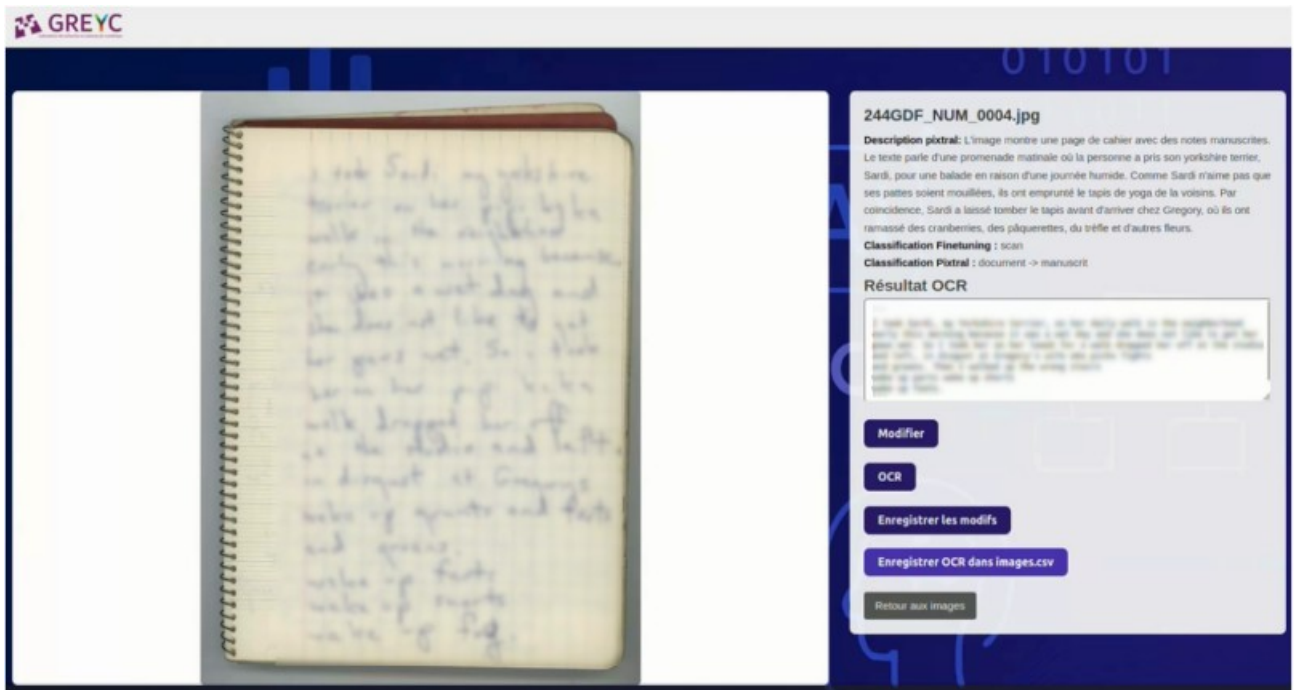

*Figure 7: Illustration du module de reconnaissance de l'écriture manuscrite dans les images de documents avec interface de retouche manuelle des erreurs résiduelles*

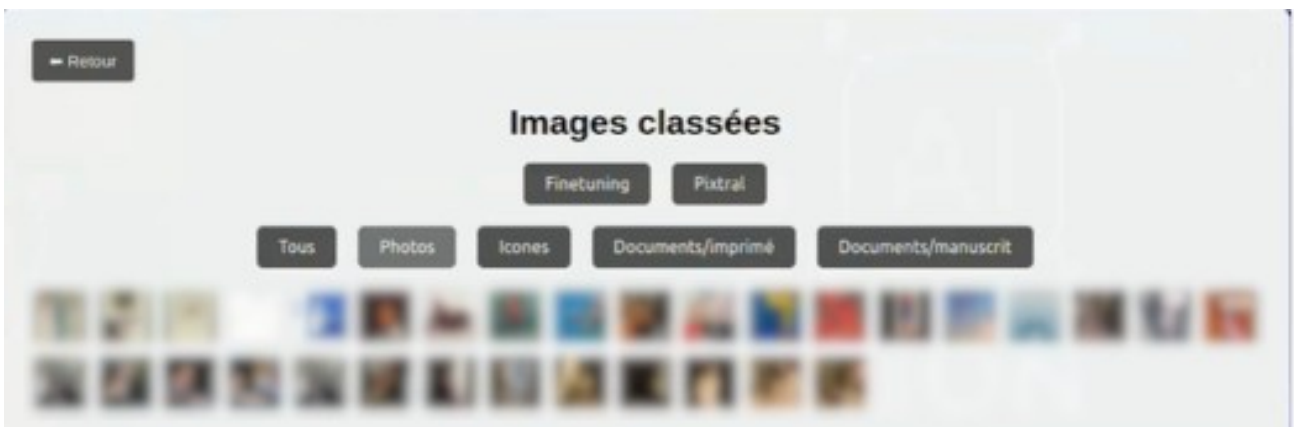

*Figure 8: Illustration du module de classification automatique des images de documents*

# 7. Détection de propos haineux et communicabilité des archives

En analysant des fonds d'archives numériques, les archivistes de l'IMEC peuvent être confrontés à des contenus parfois sensibles, pouvant parfois même heurtés. S'il semble évident que certains contenus ne devraient pas être communiqués ou divulgués, il paraît tout aussi important de prévenir les archivistes de leur présence avant qu'ils n'y soient exposés de manière impromptue, afin de limiter l'impact psychologique qui pourrait résulter d'une confrontation imprévue.

La détection des contenus sensibles n'est pas sans lien avec la question de la communicabilité de ces contenus au grand public ou aux chercheurs étudiant les fonds d'auteurs. La notion de communicabilité des contenus est en fait plus large et recouvre des notions à la fois réglementaires, juridiques, éthiques et contractuelles. La communicabilité n'est par essence pas un concept détectable de manière totalement automatique. Pour l'IMEC, qui traite des archives privées, la communicabilité d'un document repose sur l'expertise des archivistes qui sont les plus à même d'appréhender le contexte juridique particulier et les rapports à l'œuvre.

Dans ce contexte, l'analyse automatique du contenu textuel peut venir en appui au travail des archivistes en identifiant en amont les propos injurieux, discriminatoires ou haineux. Tel a été l'objectif de ce projet : réaliser un outil de détection de propos haineux pour contribuer à la gestion éthique et sécurisée des archives. 11 catégories de contenus sensibles ont été considérées : propos injurieux, propos haineux, racisme, religion, sexisme, lgbtq-phobe, handicap, xénophobie, politique, violence, menace. Cette taxonomie présente quelques chevauchements inter-catégoriels qui pourraient permettre de capter des propos non rattachés à une catégorie précise.

Pour tenter d'atteindre ce but, plusieurs étapes de constitution d'un corpus ont été mises en place : extraction et préparation de textes issus des fonds, constitution de jeux de données francophones et anglophones spécialisés dans la détection de discours haineux, enrichissement des corpus par traduction automatique, génération contrôlée via des modèles de langage pour équilibrer les catégories minoritaires. Deux modèles basés sur les transformeurs ont été considérés : CamemBERT, adapté au français, et XLM-RoBERTa, modèle multilingue robuste. Ces modèles ont été affinés en classification multi-label sur les 11 catégories de contenus sensibles.

Ce travail a permis de mettre en place une chaîne complète allant de la préparation des données à l'évaluation de modèles transformer pour la détection de propos haineux dans les archives. Il n'a cependant pas permis d'aboutir à la production d'un système exploitable, la qualité des résultats étant encore impactées par le manque de données pour entraîner les modèles.

**Référence :**

Hamza OUAZZANI CHAHDI - 4 mois. Investigation numérique d'archives d'écrivains contemporains. Communicabilité des archives des écrivains contemporains. Stage de 2A encadré par Christophe Rosenberger. Du 15 avril au 15 août 2025. [Internship Report]

**Illustration :**

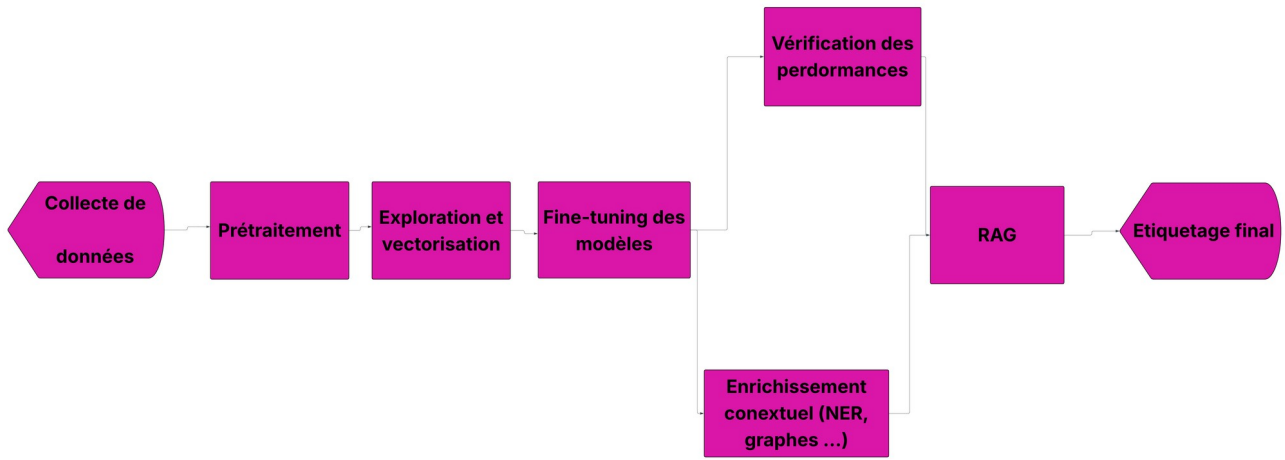

*Figure 9: Illustration du pipeline permettant la détection de propos haineux*

# BILAN

Les outils conçus au GREYC dans le cadre du projet IANEC ont tous été testés sur des données archivistiques de l'IMEC, et sont en mesure d'assister efficacement le travail archivistique de fouille numérique. On peut souligner en particulier :

- Un outil d'émulation multi-systèmes d'exploitation Mac, déjà utilisé à plusieurs reprises par l'IMEC, permet désormais de récupérer des fichiers issus de formats anciens sans avoir à gérer la complexité de configuration de ces environnements.

- Une solution permettant de rendre directement accessibles le contenu de certaines images-disques (HFS), sans passer par des solutions ad-hoc, afin de simplifier et d'automatiser le traitement massif de ces images.

- Les travaux sur la classification par typologie documentaire, qui ont été testés sur plusieurs fonds avec succès, et sont utilisables aisément dans un outil sur navigateur. Ces derniers peuvent assister le travail de l'archiviste sur l'identification initiale pour le classement, comme sur l'identification d'éléments de contenu non communicables ;

- La classification appliquée à la catégorie « image de documents », qui a été testée sur des corpus de l'IMEC, et qui permet d'envisager l'extraction automatique d'informations d'identification sur des gros corpus, notamment grâce à l'OCR qui couvre aussi les documents manuscrits, et la génération d'une description automatique du contenu.

Les perspectives à ce travail sont nombreuses. Dans un premier temps, il s'agirait de réaliser une meilleure intégration des livrables du projet à la chaîne de traitement des archives à l'IMEC afin de permettre la validation des travaux sur une plus grande diversité de fonds, mais également la production de résultats conformes aux normes archivistiques.

Les travaux menés sur la classification documentaire, l'extraction de contenu ou le résumé de documents à base de grands modèles de langues et d'intelligence artificielle se sont montrés très prometteurs et permette,t aujourd'hui d'envisager de nouvelles manières d'appréhender le traitement des archives. Cela nécessiterait cependant de consolider les infrastructures matérielles de l'IMEC pour tirer pleinement partie de ce potentiel. Il conviendrait par ailleurs de réfléchir davantage au changement méthodologique que cela pourrait engendrer sur le métier.

# STAGES ENCADRES AU GREYC

## 8. Stages encadrés en 2024

Arthur LOMBARDO - 2 mois. Amélioration de l'interface utilisateur de la plateforme d'investigation numérique G'DIP. Stage de M1 encadré par Emmanuel Giguet et Tanguy Gernot. Du 27 mai au 19 juillet 2024.

Hamid BEN OMAR - 4 mois. Intégration de modules d'investigation G'DIP à Autopsy et Préparatifs à l'investigation numérique sur d'anciens Mac OS. Stage de 2A encadré par Emmanuel Giguet et Tanguy Gernot. Du 15 avril au 19 août 2024 2024. [Internship Report]

Thomas VARIN - 3 mois. Détection et reconnaissance de visage dans des images d'archive. Stage de 2A encadré par Emmanuel Giguet et Tanguy Gernot. Du 3 juin au 23 août 2024. [Internship Report]

## 9. Stages encadrés en 2025

Mohammed Salah AISSAOUI - 4 mois. Investigation numérique d'archives d'écrivains contemporains. Classification automatique de documents. Stage de M1 encadré par Tanguy Gernot, Emmanuel Giguet et Christophe Rosenberger. Du 3 mars au 4 juillet 2025. [Internship Report]

Matthias DAVID - 4 mois. Investigation numérique d'archives d'écrivains contemporains : Fournir un service numérique innovant à l'Institut mémoires de l'édition contemporaine. Stage de M1 encadré par Tanguy Gernot, Emmanuel Giguet et Christophe Rosenberger. Du 3 mars au 4 juillet 2025. [Internship Report]

Guillaume HAUTOT - 4 mois. Investigation numérique d'archives d'écrivains contemporains. Stage de M1 encadré par Tanguy Gernot, Emmanuel Giguet et Christophe Rosenberger. Du 3 mars au 4 juillet 2025. [Internship Report]

Titouan LE BRET - 4 mois. Investigation numérique d'archives d'écrivains contemporains : Analyse forensique et traitement d'archives numériques. Stage de M1 encadré par Tanguy Gernot, Emmanuel Giguet et Christophe Rosenberger. Du 3 mars au 4 juillet 2025. [Internship Report]

Hamza OUAZZANI CHAHDI - 4 mois. Investigation numérique d'archives d'écrivains contemporains. Communicabilité des archives des écrivains contemporains. Stage de 2A encadré par Christophe Rosenberger. Du 15 avril au 15 août 2025. [Internship Report]

Dihia SLIMANA - 4 mois. Investigation numérique d'archives d'écrivains contemporains. Classification automatique de documents. Stage de M1 encadré par Tanguy Gernot, Emmanuel Giguet et Christophe Rosenberger. Du 3 mars au 4 juillet 2025. [Internship Report]